\documentclass[aps,prl,reprint,superscriptaddress]{revtex4-1}

\usepackage{lipsum}  
\usepackage{graphicx}
\usepackage{dcolumn}
\usepackage{bm}
\usepackage{hyperref}
\usepackage[mathlines]{lineno}
\usepackage{siunitx}
\usepackage[utf8]{inputenc}

\begin{document}


\title{Photodynamics of quantum emitters in hexagonal boron nitride revealed by low temperature spectroscopy}


\author{Bernd Sontheimer}
\email[]{bernd.sontheimer@physik.hu-berlin.de}
\homepage[]{www.physik.hu-berlin.de/nano}
\author{Merle Braun}
\author{Niko Nikolay}
\author{Nikola Sadzak}
\affiliation{Institut f\"ur Physik, Humboldt-Universit\"at zu Berlin, Berlin, Germany}
\author{Igor Aharonovich}
\affiliation{School of Mathematical and Physical Sciences, University of Technology Sydney, Ultimo, NSW 2007, Australia}
\author{Oliver Benson}
\affiliation{Institut f\"ur Physik, Humboldt-Universit\"at zu Berlin, Berlin, Germany}



\date{\today}

\begin{abstract}
Quantum emitters in hexagonal boron nitride (hBN) have recently emerged as promising bright single photon sources. In this letter we investigate in details their optical properties at cryogenic temperatures. In particular, we perform temperature resolved photoluminescence studies and measure photon coherence times from the hBN emitters. The obtained value of \SI{81\pm1}{\pico\second} translates to a width of $\sim$\SI{12}{\giga\hertz} which is higher than the Fourier transform limited value of $\sim$\SI{32}{\mega\hertz}. To account for the photodynamics of the emitter, we perform ultrafast spectral diffusion measurements that partially account for the coherence times. Our results provide important insight into the relaxation processes in quantum emitters in hBN which is mandatory to evaluate their applicability for quantum information processing.
\end{abstract}

\pacs{}

\maketitle

Single photon sources (SPSs) are prime candidates for myriad applications in integrated quantum photonics, quantum optics and information processing\cite{kimble_quantum_2008, lodahl_interfacing_2015, ladd_quantum_2010, aharonovich_solid-state_2016}. Fluorescent atomic defects in solids (atom-like emitters) are particularly attractive in this regard as they offer exciting opportunities for scalable quantum networks. Several systems have been investigated in details, including rare earth ions in solids\cite{zhong_nanophotonic_2015, kolesov_optical_2012}, defects in silicon carbide (SiC)\cite{castelletto_silicon_2014, kraus_room-temperature_2014, koehl_room_2011} defect centers in diamond such as the nitrogen vacancy (NV)\cite{bernien_heralded_2013, childress_diamond_2013}, the silicon vacancy (SiV)\cite{sipahigil_integrated_2016, becker_ultrafast_2016, muller_optical_2014, zhou_coherent_2017} and more recently the germanium vacancy (GeV)\cite{iwasaki_germanium-vacancy_2015, ekimov_germaniumvacancy_2015, siyushev_optical_2016, bhaskar_quantum_2016}.
Latterly, a new family of SPSs emerged in hexagonal boron nitride (hBN)\cite{tran_quantum_2016, tran_robust_2016, jungwirth_temperature_2016, chejanovsky_structural_2016, bourrellier_bright_2016, exarhos_optical_2017, schell_coupling_2017}. hBN is a wide bandgap material (ca. \SI{6}{e\volt}) and can therefore host fluorescent defect centers that can be triggered at room temperature using a sub-bandgap excitation\cite{li_atomically_2016}. While the origin of the emitters is still under debate, it is tentatively associated with the antisite nitrogen vacancy (N$_\text B$V$_\text N$)\cite{tran_robust_2016}. At room temperature these emitters exhibit remarkable properties including high brightness, polarization and short excited state lifetime, making them promising candidates for quantum technologies.

In this letter we investigate in details the optical properties of these defects at cryogenic temperatures to understand the electron phonon processes and the dephasing mechanisms. In particular, we study the temperature dependence of the properties of the spectrally narrow zero-phonon line (ZPL) from a single defect in hBN. First, we cool down the selected emitter and observe the change in linewidth and central position. At cryogenic temperature we then perform lifetime and first order coherence measurements to further quantify optical dephasing mechanisms. We find the time scale of an apparent spectral diffusion process that cause the line to be spectrally much broader then natural linewidth.

\begin{figure}
\includegraphics[width=\columnwidth]{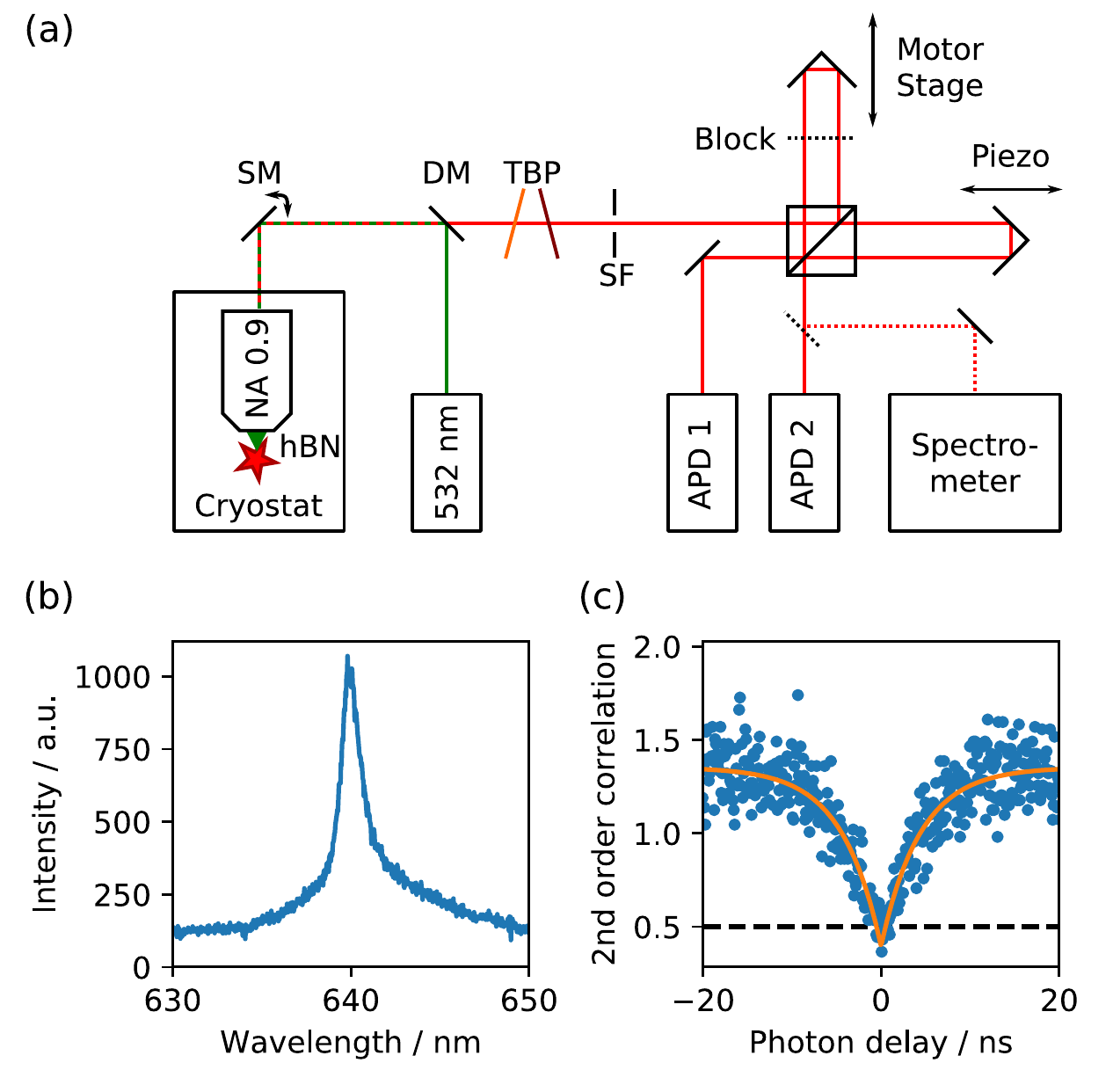}%
\caption{Setups and basic characterization: (a) Schematic of the experimental setup: SM: x/y-scanning mirrors, DM: dichroic mirror, TBP: tunable bandpass consisting of a long and a short pass filter, SF: spacial, confocal filtering system. By blocking one interferometer arm the detection path effectively becomes a typical Hanbury Brown and Twiss configuration for 2nd order correlation measurements. One output of the beamsplitter can be directed to the spectrometer, allowing for convenient emitter tracking during spectral measurements. (b) Room temperature spectrum of a single photon emitter in hBN.  (c) Close-up of the $g^{(2)}$-function around $\tau=0$ shows typical single photon anti-bunching behavior with $g^{(2)}(0)<0.5$. (blue: data, orange: fit, black dashed: $g^{(2)}(\tau)=0.5 $)\label{fig1}}
\end{figure}

We conduct the experiments with a home-built confocal microscope, sketched in Fig.~\ref{fig1}(a), to selectively excite defects in hBN flakes. The sample preparation was adapted form \cite{tran_robust_2016}. In the setup, collected photons are guided through a Michelson interferometer which allows for measurements of the photon coherence time $\tau_c$ as well as the spectral diffusion time $\tau_d$. By blocking one of the interferometer arms, the detection path is effectively converted to a Hanbury Brown and Twiss configuration for standard second order correlation measurements (correlation electronics: PicoHarp300, PicoQuant). By redirecting the detection path from one of the avalanche photo diodes (APDs, Count20, Laser Component) to a spectrometer (SP500i, Princeton Instruments), spectra can be measured while conveniently locking on the emitter intensity with the other APD. The emitters themselves, as well as the objective lens (NA 0.9, Mitutoyo) are placed inside a flow cryostat (CryoVac), allowing for highly efficient photon collection at temperatures between \SI{300}{\kelvin} and \SI{5}{\kelvin}. At cryogenic temperatures a pair of widely tunable long and short pass filters (704 nm VersaChrome Edge, Semrock) acts as a $\sim$\SI{5}{\nano\meter} bandpass centered at the ZPL of the selected single emitter, to maximize the signal to background ratio. An additoinal \SI{620}{\nano\meter} long pass filter and the dichroic mirror further suppress the \SI{532}{\nano\meter} excitation diode laser. Lifetime measurements were performed using a picosecond pulsed \SI{532}{\nano\meter} laser (LDH-P-FA-530, PicoQuant). All of the following measurements were performed on the same specific defect center over the course of two days. The emitter was chosen because of its already narrow and stable ZPL at room temperature and its clear anti-bunching dip shown in Fig.~\ref{fig1}(b) and Fig.~\ref{fig1}(c).

\begin{figure}
\includegraphics[width=\columnwidth]{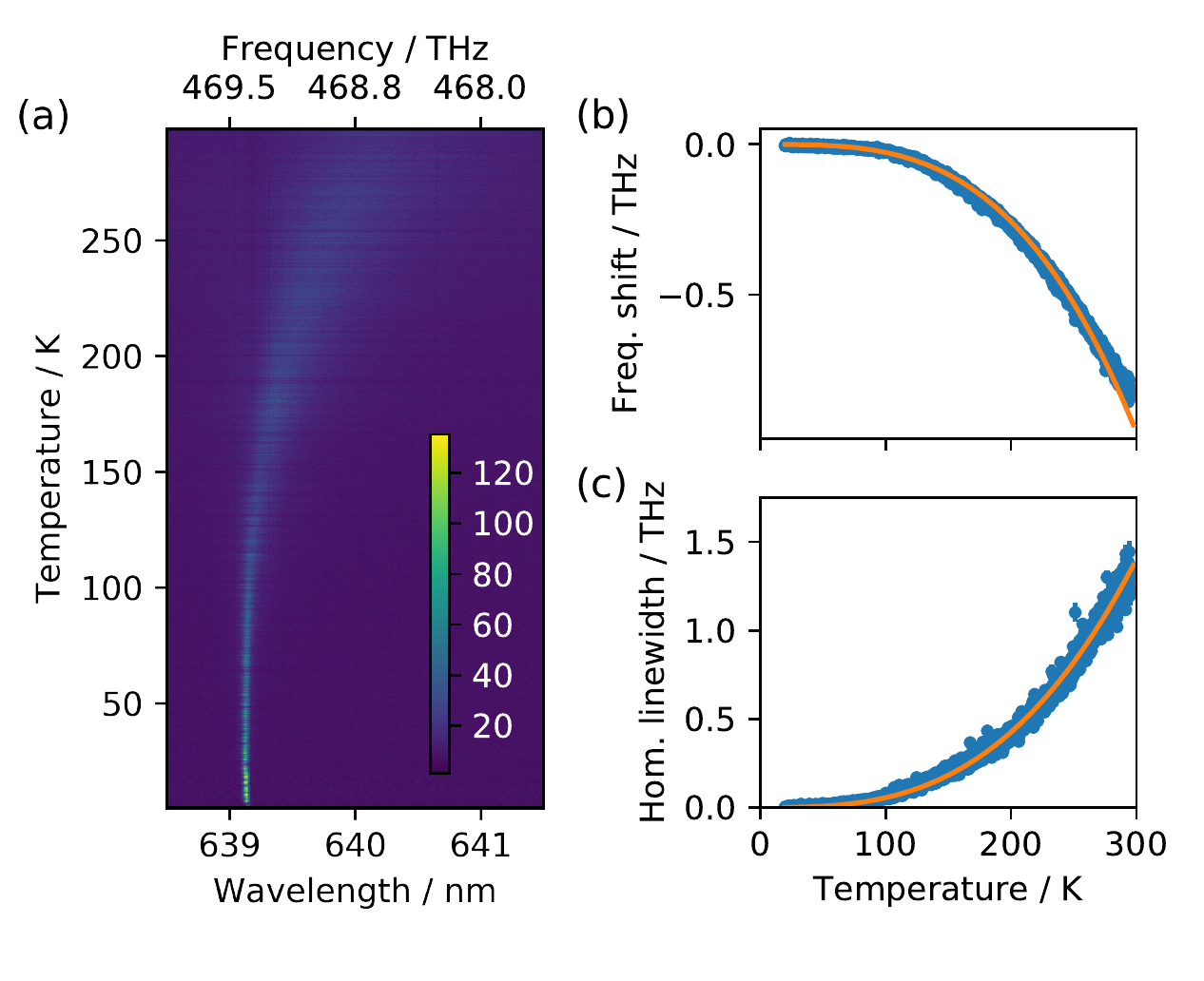}%
\caption{Temperature-dependent measurements: (a) Spectra,  (b) frequency shift and (c) homogeneous linewidth of the ZPL. The values in (b) and (c) are extracted from fits to a Lorentzian model following the approach from \cite{neu_low-temperature_2013}. The solid lines in (b) and (c) are fits to a power law, resulting in a $T^{3.39(1)}$ and $T^{2.94(1)}$ dependency respectively.\label{fig2}}
\end{figure}

Fig.~\ref{fig2}(a) shows a heat map of the spectra recorded during the cooling process. Each individual spectrum was integrated for \SI{500}{\milli\second}. The temperature resolution was limited to \SI{0.1}{\kelvin} and spectra within each temperature step were averaged, in order to obtain an optimal signal to noise ratio. Cooling from room temperature to \SI{5}{\kelvin} took about \SI{1}{\hour} and the cooling rate was kept constant. Down to a temperature of \SI{20}{\kelvin} the lineshape and position of the ZPL significantly change. To further invesitage on those changes we follow the approach given in \cite{neu_low-temperature_2013}: In general, the ZPL shape is governed by homogeneous and inhomogeneous broadening mechanisms, which lead to a Lorentzian and Gaussian shape respectively. For defect centers in solid states the predominant mechanisms are spectral diffusion (inhomogeneous) and phonon broadening (homogeneous)\cite{wolters_measurement_2013, neu_low-temperature_2013, muller_phonon-induced_2012, fu_observation_2009}. If both occur, the resulting line shape is given by the convolution of both - a Voigt profile:
\begin{equation}
  I_V(\nu) = \frac{A \textrm{Re}[w(z)]}{\sigma\sqrt{ 2 \pi}}
\end{equation}
with 
\begin{equation}
  w(z) = e^{-z^2}\text{erfc}(-\mathrm iz) \text{\ \ \ \ and \ \ \ \ }   z = \frac{\nu-\mu +\mathrm i\gamma}{\sigma\sqrt{2}}
\end{equation}
where $A$ is the peak amplitude, $\mu$ is the center, $f_G=2\sigma\sqrt{2\ln 2}$ is the full width half maximum of the Gaussian contribution and $f_L=2\gamma$ is the full width half maximum of the Lorentzian contribution. In the following discussion, the obtained linewidths were corrected for the spectrometer response function, which also follows a Voigt profile (see supplemental material).
In agreement with other defect centers in broad bandgap semiconductors\cite{muller_phonon-induced_2012, neu_low-temperature_2013} we see a transition from a predominantly Gaussian line shape to a Lorentzian-like spectrum for increasing temperatures. In analogy to defects in nano diamond\cite{neu_low-temperature_2013, wolters_measurement_2013}, this indicates, that the phonon broadening increases with temperature while the spectral diffusion stays mostly temperature independent. In order to fit the measured spectra to theoretical Voigt profiles we adapt the procedure established for silicon vacancy centers in diamond described in \cite{neu_low-temperature_2013}: at low temperatures the Lorentzian width of the measured Voigt profiles is assumed to be far below the spectrometer resolution and therefore is fixed to the value obtained from the spectrometer response function $f_{L,\text{spec.}}=\SI{3.7}{\giga\hertz}$. For up to \SI{20}{\kelvin} fits to the measured profiles yield a constant value of $f_G=\SI{21.3}{\giga\hertz}$. This is the temperature independent spectral diffusion linewidth. For higher temperatures the Gaussian width is fixed to this value and $f_L$ is left as a unrestrained fit parameter. The graphs in Fig.~\ref{fig2}(b) and (c) show the a shift of the central frequency and a broadening of the homogeneous linewidth extracted from the resulting fits for all temperatures. The frequency redshift follows a power law proportional to $T^{3.39(1)}$. This compares well with measurements on SiV and Chromium defect centers in diamond, where a $T^3$ dependent shift is commonly attributed to fluctuating fields which are created as a phonon modulates the distance between the color center and other impurities in the crystal\cite{neu_low-temperature_2013, muller_phonon-induced_2012}. For the linewidth we observe a broadening with a $T^{2.94(1)}$ power law for increasing temperatures. Again, this is similar to the behavior found in the diamond defects mentioned above. Moreover, a power law close to $T^3$ for the linewidth was observed in solid state systems exhibiting significant inhomogeneous ZPL broadening\cite{wei_spectral_1991, krustok_temperature_2006} where it was also linked to the influence of impurities in the host material\cite{hizhnyakov_zero-phonon_2002}. This corresponds well with our results, since the two dimensional nature of hBN implies a rather unprotected environment for the embedded defect center.

\begin{figure}
\includegraphics[width=\columnwidth]{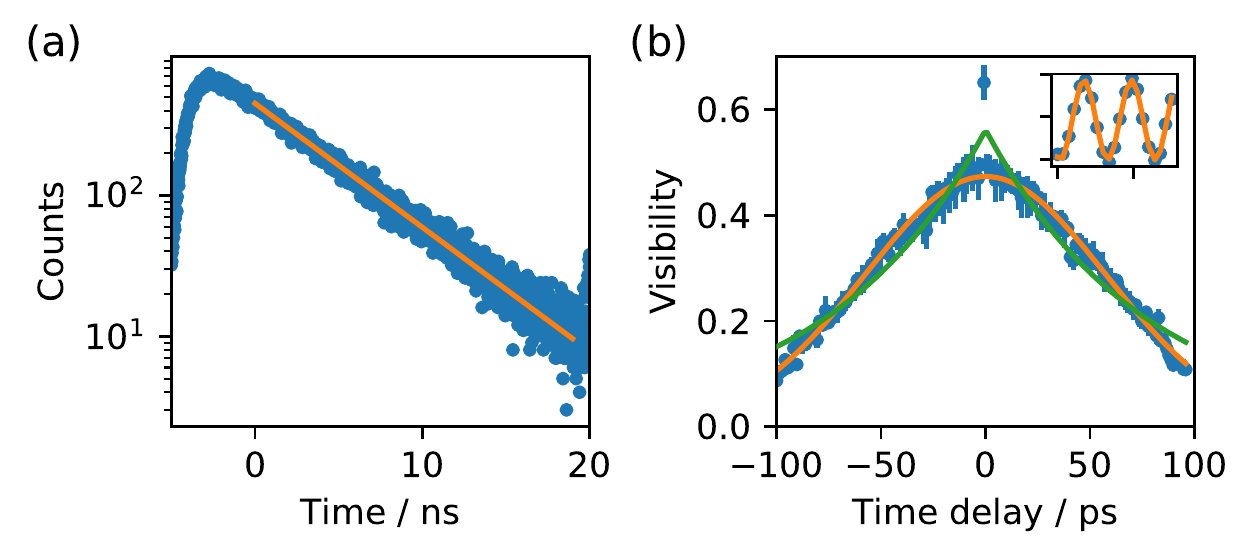}%
\caption{Lifetime and coherence measurements of a single defect in hBN: (a) Lifetime measurement (blue: data, orange: exponential fit with $\tau= \SI{4.93 \pm 0.02}{\nano\second} $ (b). Measured first order coherence measurement (blue: data, green: exponential fit with $\tau_{c,e}=\SI{76\pm2}{\pico\second}$, orange: Gaussian fit with $\tau_{c,G}=\SI{81\pm1}{\pico\second}$): Each visibility data point is the result of a fit to an piezo scan of about two
 fringes during ca. \SI{250}{\milli\second} at a fixed stage position (inset).\label{fig3}}
\end{figure}

Next, we explore the actual limit of the single hBN defect at low temperature. A measurement of the excited state lifetime $\tau$ [Fig.~\ref{fig2}(a)] reveals a natural linewidth  $f_n=(2\pi\tau)^{-1}\approx\SI{32}{\mega\hertz}$. This ultimate limit is far below the spectrometer resolution. Therefore, we performed interferometric measurements to directly determine the real linewidth. This is done by first order coherence measurements on the ZPL at \SI{5}{\kelvin} with the Michelson interferometer illustrated in Fig.~\ref{fig1}(a). Fig.~\ref{fig3}(b) shows the obtained interference visibility as a function of the relative time delay due to the interferometer path length difference. For each data point the motor stage was moved to a specific delay position and the interferogram was measured locally by scanning the piezo over around two interference fringes (Fig.~\ref{fig3}(b) inset). The visibility was then extracted by fitting a sine function with an offset to the interferogram and calculating the absolute value of the quotient of the resulting oscillation amplitude and the offset. From Fig.~\ref{fig3}(b) we identify a Gaussian-like visibility profile centered at the position of equal interferometer arm lengths. This is in very good aggreement with the predominantly Gaussian emitter line shape at cryogenic temperatures seen with the spectrometer \cite{salamon_michelson_1974}. A Lorentzian shaped spectral line would result in an exponentially decaying visibility profile.

The coherence time of \SI{81\pm1}{\pico\second} corresponds to a spectral width of $f_c=1/\tau_c\approx \SI{12}{\giga\hertz}$; roughly 40 times wider than the Fourier limited linewidth. This measurement represents to best of our knowledge the first linewidth measurement from a single emitter in a two dimensional material and specially hBN. Although this is still far from the Fourier limit, we would like to point out that the value is comparable or even better than the ones obtained from measurements on chromium or NV centers in diamond\cite{muller_phonon-induced_2012, jelezko_coherence_2003}.

\begin{figure}
\includegraphics[width=\columnwidth]{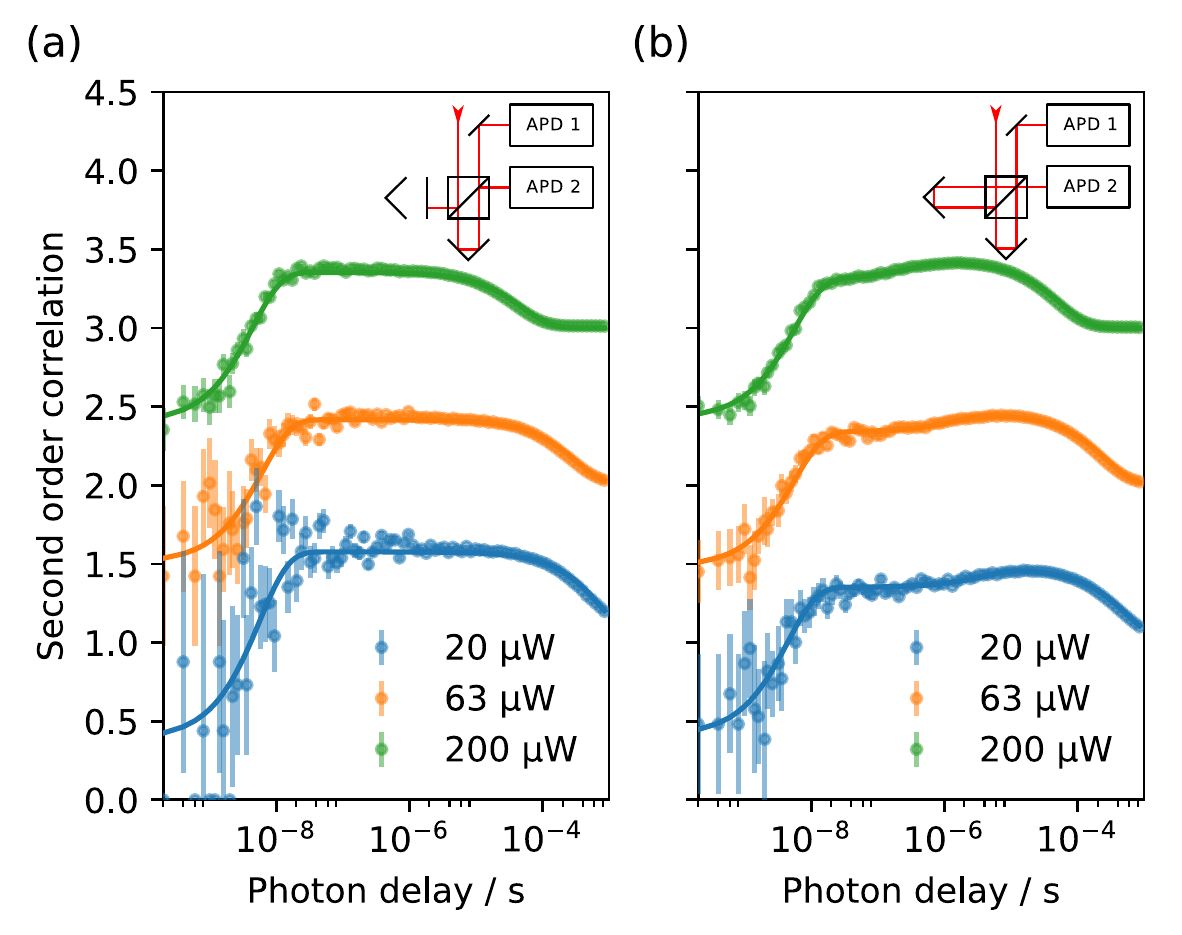}%
\caption{Second order correlations of the emitter signal for different excitation laser powers with (a) detection in a typical Hanbury Brown and Twiss configuration and (b) detection at the output ports of a Michelson interferometer. While (a) follows the $g^{(2)}$-function of a three level system, (b) additionally shows the effect of spectral diffusion (points with error bars: correlated data, solid lines: fits, the orange and green curves have an offset of $+1$ and $+2$ respectively). \label{fig4}}
\end{figure}

As seen above the reason for the large difference between the lifetime limited width and the one obtained with the spectrometer and interferometer is inhomogeneous broadening. In the following we investigate further on the underlying mechanism: We propose spectral jumps. Many of those jumps occuring during the data aquisition (typically \SI{250}{\milli\second}) will broaden the intrinsically lifetime-limited linewidth. We can confirm the presence and determine the time scale of spectral jumps with second order correlation measurements following the approach demonstrated in \cite{wolters_measurement_2013}. Photons emitted from a hBN defect follow the $g^{(2)}$-function of a three level system
\begin{equation}
  g^{(2)}(\tau) = 1 - (1+A)\cdot\text{e}^{-\frac{\tau}{t_1}} + A \cdot\text{e}^{ -\frac{\tau}{t_2}}
\end{equation}
with the bunching amplitude $A$, the anti-bunching time $t_1$ and the bunching time $t_2$. In the experiment we measure the $g^{(2)}$-function by blocking one arm of the Michelson interferometer (see Fig.~\ref{fig4}(a) inset) and correlating the APD signal. This resembles a typical Hanbury Brown Twiss configuration. The treatment of additional background with a Poissonian photon statistics is described in the supplement. With the interferometer arm unblocked, the measurement is additionally modified by the presence of the interferometer, which transforms frequency modulations, such as spectral diffusion, into intensity changes on the APDs. In this way even very fast spectral diffusion or spectral jumps can be measured\cite{wolters_measurement_2013}. If the path length difference between the two arms of the interferometer and therefore the fringe width is smaller than the spectral diffusion width, the resulting  $g^{(2)}_{LR}$-function reads as:
\begin{equation}
  g^{(2)}_{LR}(\tau) =  \left(1 - \frac{c^2}{2}\cdot\text{e}^{\frac{\tau}{t_d}}\right)\cdot g^{(2)}(\tau)
\end{equation}
where $c$ is the interferometric contrast and $t_d$ the spectral diffusion time\cite{wolters_measurement_2013}. Fig.~\ref{fig4} shows the resulting second order correlations for three different laser powers. Table~\ref{tab1} summarizes the relevant time parameters of the fits. For increasing powers, the anti-bunching time $t_1$ remains nearly unchanged at about \SI{5}{\nano\second} while the bunching time $t_2$ significantly reduces from \SI{814\pm3}{\micro\second} to \SI{49.73\pm0.07}{\micro\second} and respectively \SI{575\pm1}{\micro\second} to \SI{56.59\pm0.03}{\micro\second} for the different configurations. This behavior is typical for three level systems. The spectral diffusion measurements shows that the time $t_d$ also decreases for increasing power from \SI{3.9\pm0.2}{\micro\second} to \SI{0.320\pm0.008}{\micro\second}. However, this is still at least two orders of magnitude slower than the lifetime of the emitter, which means the defect emitts multiple consecutive photons before it spectrally jumps.
\begin{table}
\caption{Selected fit parameters for fits in Fig.~\ref{fig4}.\label{tab1}}
\begin{ruledtabular}
\begin{tabular}{l c c c c}
& Power / \si{\micro\watt} & $t_1$ / \si{\nano\second} & $t_2$ / \si{\micro\second} & $t_d$ / \si{\micro\second}\\
\hline
          & $20$  & $5    \pm 1$    & $814   \pm 3$   &  - \\
$g^{(2)}$     & $63$  & $6.2  \pm 0.8$  & $321.3 \pm 0.6$ &  - \\
          &$200$  & $4.4  \pm 0.2$  & $49.73 \pm 0.07$&  -\\
\hline
          & $20$  & $4.4  \pm 0.7$  & $575.0 \pm 1.0$ & $3.9 \pm 0.2$\\ 
$g^{(2)}_{LR}$& $63$  & $5.5  \pm 0.3$  & $268.4 \pm 0.2$ & $1.4 \pm 0.5$\\
          &$200$  & $5.4  \pm 0.1$  & $56.59 \pm 0.03$& $0.320 \pm 0.008$\\
\end{tabular}
\end{ruledtabular}
\end{table}

The presence of the spectral diffusion is in accord with the potential defect structure of the nitrogen antisite that has a permanent dipole moment\cite{tran_robust_2016}. The rather fast values may suggest charging and discharging of the defect that will result in a small wavelength shift due to the DC Stark effect. The extra charge may originate from the rich nitrogen environment within the hBN lattice.
Moreover, the presence of ultrafast spectral diffusion verifies again that the phonon coupling is not the dominant process causing the low coherence times. We also note that the recent development of engineering the hBN single emitters in large exfoliated materials can yield a more robust and clean environment that potentially reduces the spectral diffusion\cite{exarhos_optical_2017, chejanovsky_structural_2016, choi_engineering_2016}. This is analogue to comparing emitters in nanodiamonds vs emitters in a bulk, that often exhibit much stable and superiror photophysical properties.
On the other hand due to the twodimensional nature of the host material the defect is highly exposed to its surroundings. We therefore suspect the spectral diffusion to be strongly dependend on its close environment and therefore potentially applicable to very localized sensing applications.

To summarize, we presented detailed characterization of an optically active point defect in hBN. We performed temperature dependent photo luminescence experiments and measured the photon coherence time of the single photon emitter. The obtained photon coherence time \SI{81\pm1}{\pico\second} is still less than the lifetime limited value. This can be explained by the ultrafast spectral diffusion present in these emitters which causes an inhomogeneous broadening of the line. The presented results are important for further understanding of the hBN defect centers and lead to new ideas for quantum information and sensing application based on twodimensional wide bandgap host materials.

\begin{acknowledgments}
We greatfully acknowldege funding from the DFG (Sfb 951) the Einstein Foundation Berlin (ActiPlAnt) and CNPq Brasil (science without borders program). We also thank for the financial support from the Australian Research Council (DE130100592) and the Asian Office of Aerospace Research and Development grant FA2386-15-1-4044. Moreover, I. A. acknowledges the generous sponsorship provided by the Alexander von Humboldt Foundation.
\end{acknowledgments}

\bibliography{main}

\end{document}



\title{Supplemental material for "Photodynamics of quantum emitters in hexagonal boron nitride revealed by low temperature spectroscopy"}




\pacs{}

\maketitle

\section{Instrument response function of the spectrometer}
The instrument response function of the spectrometer was determined by exchangeing the excitation laser with a diode laser at \SI{638}{\nano\meter} (Velocity TLB-6304-H, Newport) and recording a spectrum of the laser reflex from the sample surface.
\begin{figure}[!h]
\includegraphics[width=\columnwidth]{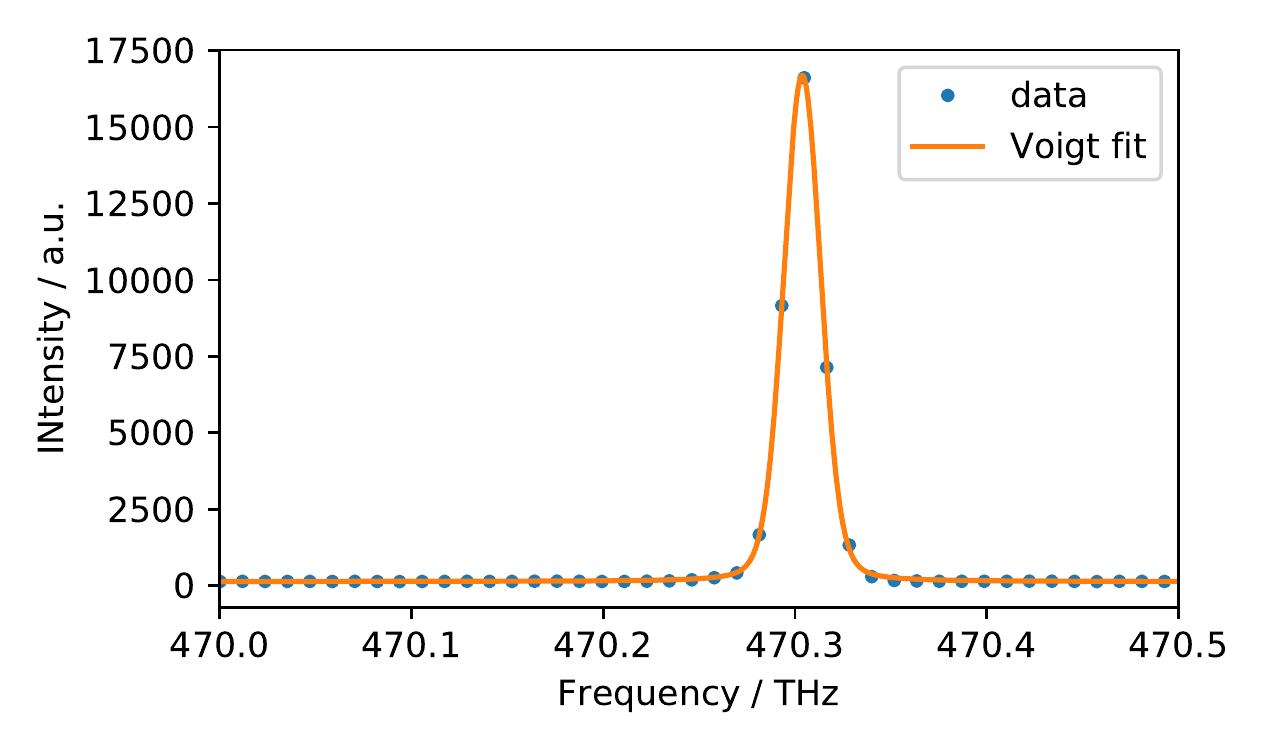}%
\caption{Recorded instument response function of the spectrometer and a Voigt fit to the data.\label{sfig1}}
\end{figure}

Fig.~\ref{sfig1} shows the recorded laser spectrum which we will take as the spectrometer response function. It follows a Voigt profile with the Lorentzian width $f_{L,spec}=\SI{3.72\pm0.10}{\giga\hertz}$ and the Gaussian width $f_{G,spec}=\SI{20.85\pm0.07}{\giga\hertz}$. Effects of the laser linewidth are be neglected since the specified width below \SI{300}{\kilo\hertz} is more than one order of magnitude lower than those values.
 
\section{Temperature dependent change of the ZPL shape}
\begin{figure}[!ht]
\includegraphics[width=\columnwidth]{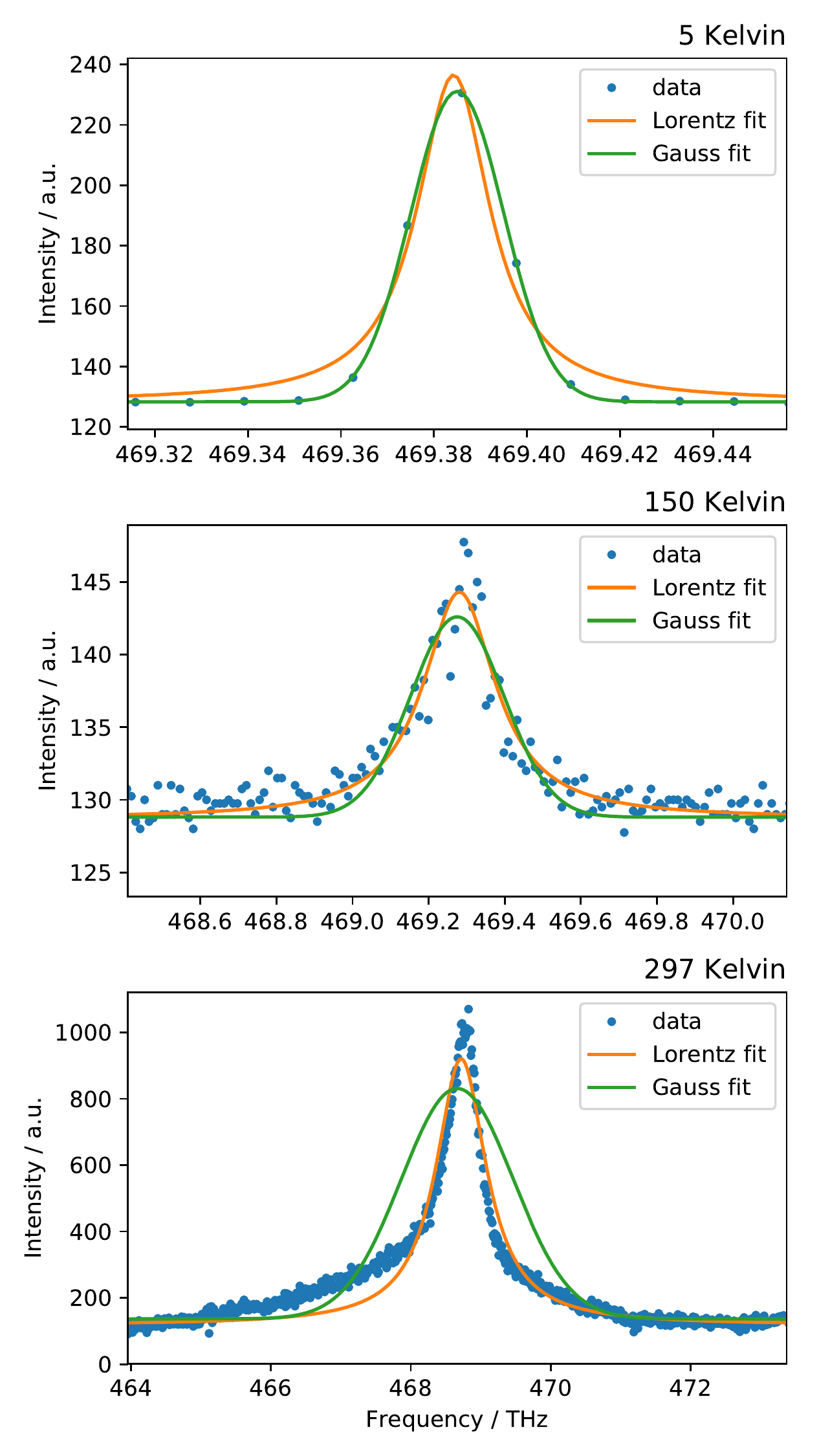}%
\caption{Visualizaiton of the temperature dependent change of the ZPL shape. All three figures show the measured ZPL spectrum at a specific temperature and a Lorentzian as well as a Gaussian fit to the data. At \SI{5}{\kelvin} a Gaussian profile matches the data well, while a Lorentzian profile does not. At \SI{150}{\kelvin} it already starts to become evident that the data is now better fitted by a Lorentzian profile. Finally, at \SI{297}{\kelvin} the situation is obvious and the ZPL is governed by a Lorentzian profile.\label{sfig2}}
\end{figure}

In Fig.~\ref{sfig2} an additional visualizaion of the transition of the ZPL shape from a mainly Gaussian profile at low temperatures to a mainly Lorentzian profile for elevated temperatures is shown. This is in accord to an increasing homogeneous broadening for higher temperatures and a temperature independend inhomogeneous broadening.

\section{Introducing background to 2nd order autocorrelation functions}

The measured $\widetilde{g}^{(2)}$ function is defined as:
\begin{equation}
    \widetilde{g}^{(2)}(\tau)=\frac {\left\langle \widetilde I(t)\widetilde I(t+\tau )\right\rangle }{\left\langle \widetilde I(t)\right\rangle ^{2}}
\label{eq:1}
\end{equation}
where $\widetilde I(t) $ is the experimentally measured signal intensity curve and $\tau$ is the time delay between two detection events. We now assume that the measured signal is the sum of the examined (single photon) emitter fluorescence and uncorrelated background $\widetilde I(t) = I(t) + B$ and insert this into Eq.~(\ref{eq:1}).
\begin{eqnarray}
    \widetilde{g}^{(2)}(\tau)&=&\frac {\left\langle (I(t)+B)\cdot(I(t+\tau)+B)\right\rangle }{\left\langle I(t) + B\right\rangle ^{2}}\\
    &=&\frac {\left\langle I(t)I(t+\tau)\right\rangle + \left\langle I(t)B\right\rangle +  \left\langle I(t+\tau)B\right\rangle  + \left\langle B^2\right\rangle }{\left\langle I(t) + B\right\rangle ^{2}}
\end{eqnarray}
After time averageing ($\left\langle \widetilde I(t)\right\rangle = I$ and $\left\langle \widetilde B\right\rangle = B$)  we can identify and isolate the pure emitter $g^{(2)}$-function in the equation
\begin{equation}
    \widetilde{g}^{(2)}(\tau)=\frac { g^{(2)}(\tau)I^2 + 2IB + B^2}{I^2+2IB+B^2}.\label{eq:2}
\end{equation}
Note that $I$ and $B$ are the average emitter signal and background noise intensities in the experiment. For the performed fits the signal to noise ratio was left as a free parameter and the adequate 2nd order correlation function was inserted into Eq.~(\ref{eq:2}).
 